\documentclass[aps,prd,twocolumn,showpacs,groupedaddress,10pt,nofootinbib]{revtex4}
\usepackage{graphicx}
\usepackage{amssymb}
\usepackage{amsmath}
\usepackage{amsfonts}
\usepackage[latin1]{inputenc}
\newcommand{\be}{\begin{equation}}
\newcommand{\ee}{\end{equation}}
\newcommand{\bea}{\begin{eqnarray}}
\newcommand{\eea}{\end{eqnarray}}

\begin{document}

\title{Phase space analysis of quintessence fields trapped in a Randall-Sundrum Braneworld: a refined study}

\author{Dagoberto Escobar}\email{dagoberto.escobar@reduc.edu.cu}
\affiliation{Departamento de F\'isica Universidad de Camagüey, Cuba.}

\author{Carlos R. Fadragas}\email{fadragas@uclv.edu.cu}
\affiliation{Departamento de F\'isica, Universidad Central de Las
Villas, Santa Clara \enskip CP 54830, Cuba.}

\author{Genly Leon}\email{genly@uclv.edu.cu}
\affiliation{Departamento de Matem\'atica, Universidad Central de Las
Villas, Santa Clara \enskip CP 54830, Cuba.}
\affiliation{Instituto de F\'{\i}sica, Pontificia Universidad de Cat\'olica de Valpara\'{\i}so, Casilla 4950, Valpara\'{\i}so, Chile.}

\author{Yoelsy Leyva}\email{yoelsy.leyva@fisica.ugto.mx}
\affiliation{Divisi\'on de Ciencias e Ingenier\'ia de la Universidad de Guanajuato, A.P. 150, 37150, Le\'on, Guanajuato, M\'exico.}

\date{\today}

\begin{abstract}

In this paper we investigate, from the dynamical systems
perspective, the evolution of an scalar field with arbitrary
potential trapped in a Randall-Sundrum's Braneworld of type II.
We consider an homogeneous and isotropic Friedmann-Robertson-Walker
(FRW) brane filled also with a perfect fluid. Center Manifold Theory is employed to obtain sufficient
conditions for the asymptotic stability of de Sitter solution. We
obtain conditions on the potential for the stability of scaling
solutions as well for the stability of the scalar-field dominated
solution. We prove the there are not late time attractors with
5D-modifications (they are saddle-like). This fact correlates with a transient primordial inflation. In the
particular case of a scalar field with potential
$V=V_{0}e^{-\chi\phi}+\Lambda$ we prove  that for $\chi<0$ the de Sitter solution is asymptotically stable. However, for $\chi>0$ the de Sitter solution is unstable (of saddle type). 

\end{abstract}

\maketitle

\section{Introduction}

The Randall-Sundrum brane of type II model (RS2), introduced originally
as an alternative mechanism to the Kaluza-Klein compactifications
\cite{Randall1999b}, have been intensively studied in the last
years, among other reasons, because its appreciable cosmological
impact in the inflationary scenario \cite{Hawkins2001a,
Huey2001, Huey2002}. The setup of the model start with the
particles of the standard model confined in a four dimensional
hypersurface with positive tension embedded in a 5-dimensional
bulk with negative cosmological constant. It is well-known that
the cosmological field equations on the brane are essentially
different from the standard 4-dimensional cosmology \cite{Binetruy2000a, Binetruy2000, Bowcock2000}. In fact the
appearance of a quadratic term  of the total energy density in the
Friedmann equation is responsible for gravitational modifications
at very high energy. The dynamics of the Universe during the quadratic dominant stage have been studied by several authors. In \cite{Maeda2001, Mizuno2003} it is shown that the (inverse) power-law potential model allow wide conditions for a successful quintessence scenario in contrast with exponential potential and k-essence models which do not present favorable scenarios. The stability of the scaling solutions for the case of power-law potential model in the presence of a perfect fluid with arbitrary barotropic index $\gamma$ was developed by \cite{Savchenko2003}. This study was extended by \cite{Tsujikawa2004} to a generalized background $H^2\propto\rho_T^n$ for an arbitrary $n$ (the RS$2$ case is recovery when $n=2$). Another interesting and recently feature of
this scenario is that the fate of the cosmic expansion can be
modified if the energy density of some matter component grow as
the expansion proceeds \cite{Garcia-Salcedo2011} \footnote{This
kind of modification do not appear if the energy density of the
matter content in the brane dilutes with the cosmic expansion as
occurs with the common matter sources: quintessence scalar field,
radiation, dust, etc.}

Several astrophysical observations such as Type Ia Supernovae
\cite{Riess2007, Davis2007, Wood-Vasey2007}, Large Scale
Structure \cite{Tegmark2004} and Cosmic Microwave Background
\cite{Jarosik2011, Larson2011, Komatsu2011} strongly confirm that
our universe currently experiences an accelerated expansion phase.
Several models, based on RS2 framework, have been proposed in
order to deal with these feature of our universe. One approach for
explaining the accelerated expansion is the Modified Chaplygin Gas
\cite{Rudra2012}. For this models was showed that the Universe
follows a power law-expansion around the critical points. Another
important approach consist in adding a self-interacting scalar
field to the matter inventory in the brane \cite{Gonzalez-Diaz2000, Maeda2001, Majumdar2001h, Nunes2002, Sami2004}. Scalar fields arise in a natural way in particle physics and they can act as a
candidate for dark energy playing the roles such that
quintessence, phantom, quintom, tachyons field etc.
\cite{Copeland2006}

The dynamical behavior of scalar field coupled with a barotropic
fluid in a spatially flat Friedmann-Robertson Walker universe has
been studied by many authors, see for instance the references
\cite{Copeland1998, Aguirregabiria2004c, Lazkoz2007, Fang:2008fw,
Leon2009, Leon2010, leon2011a}. A natural generalization of
\cite{Copeland1998}, is to include higher-dimensional behavior
(RS2 scenario). This program was carried out in \cite{Copeland2005} where was found the corresponding scalar field potentials which lead to attractor scaling solution for several energy density modifications to the Friedmann equation; for the RS2 framework the potential $V\propto \text{cosech}^2(A \phi)$ was found \footnote{At early times, where the quadratic energy term dominates, this potential behave as a (inverse) power-law potential being consistent with previous analysis \cite{Maeda2001, Mizuno2003}.}. The dynamics of a scalar field with
constant and exponential potentials was investigated in \cite{Gonzalez2009}. These results were extended
to a wider class of self-interaction potential in \cite{Leyva2009}
using a method proposed by \cite{Fang:2008fw} supporting the idea
that this scenario modifies gravity only at very high energy/short
scales (UV modifications only) having an appreciable impact on
primordial inflation but does not affecting the late-time dynamics
of the Universe \footnote{A word of caution: this claim is not in
general true, specially if the energy density of the matter
trapped in the brane increase at late times
\cite{Garcia-Salcedo2011}}. In this paper we make a step forward
with respect to the previous studies by exploring more deeply the
dynamics in the phase space associated to this scenario around
both hyperbolic an non-hyperbolic critical points. The last
subject cannot be consistently studied with the help of linear
analysis, but using the Center Manifold Theory. Our claim is that
the more interesting solution are the non-hyperbolic critical
points, in particular the de Sitter critical points.

In this paper we employ the Center Manifold Theory to obtain
sufficient conditions for the asymptotic stability of de Sitter
solution and for proving that here are not late time attractors
with 5D-modifications. This fact correlates with a transient
primordial inflation. Also we provide conditions on the potential
for the stability of scaling solutions as well for the stability
of the scalar-field dominated solution.

The paper is organized as follows. In Section \ref{sIII} we give
the essential details of the Randall-Sundrum Model and deals with the
dynamical system analysis, these include the center manifold
study. We explore the dynamics of an scalar field with exponential
potential plus a Cosmological Constant trapped on the brane using
the previous results in Section \ref{sIV}. Section \ref{sV} is
devoted to the physical discussion of the above results, while the
conclusions are given in Section \ref{sV}.

\section{Dynamical systems analysis of the FRW brane} \label{sIII}

In this section we will focus our attention in a brane-world model
where an scalar field, with arbitrary self-interaction potential,
is trapped on a RS2 brane. In the flat FRW metric, the field
equations read \cite{Langlois2003d, Brax2003d,
Langlois2004b, Maartens2004}:

\begin{align}
&H^{2}= \frac{1}{3}\rho_{T}\left(1+ \frac{\rho_{T}}{2\lambda}\right)+\frac{2{\cal U}}{\lambda} \label{EqF}\\
&2\dot H=-(1+\frac{\rho_T}{\lambda})(\dot\phi^2+\gamma\rho_m)-\frac{4{\cal U}}{\lambda}\label{EqR}\\
&\dot\rho_m=-3\gamma H\rho_m \label{EqC}\\
&\ddot\phi+\partial_\phi V=-3H\,\dot\phi\label{EqK} \end{align}
where we have used the Randall-Sundrum fine tunning condition i.e., we neglect the
cosmological constant term ($\Lambda_4=0$). $\rho_{T}=\rho_{\phi}+\rho_{m}$, $\lambda$ is the brane
tension, $\gamma$ is the barotropic index of the background fluid,
$V$ is the scalar field self-interaction potential. $\cal U$$(t)=\frac{C}{a(t)^4}$ is the dark radiation term which arises from a non-vanishing bulk Weyl tensor being $C$ a constant parameter related with black hole mass in the bulk: if the bulk is AdS$-$Schwarzschild $C\neq 0$ \cite{Brax2003d}. When the black hole mass vanishes, the bulk geometry reduces to AdS, and $C=0$ \cite{Bowcock2000, Maartens2004}. In the following we will study the latter case i. e., we will not consider here the dark radiation term . Here and
throughout, we use $\partial_{\phi}$ to denote derivative with
respect to $\phi.$

From the Friedmann equation (\ref{EqF}) it is
deduced how the brane effects modify the early time dynamics: at
high energy ($\rho_{T}>>\lambda$) this equation reduces to
$H\propto \rho_{T}$. A late times, due to the expansion rate, the
energy density of the matter trapped in the brane dilutes
($\rho_{T}<<\lambda$), and the standard $4$D TGR behavior is
recovered, leading to $H\propto \sqrt{\rho_{T}}$.

Having presented the cosmological equations, our purpose now is to
define a dynamical system from \eqref{EqF}-\eqref{EqK} in order to
examine all possible cosmological behaviors. As we know dynamical
systems techniques provides one of the better tools for obtaining
useful information about the evolution of a wide class of
cosmological models \footnote{See, for instance, the seminal work
\cite{Copeland1998}, and \cite{coley2003b, Tavakol2005B}.}. In
order to take advantage from these tools, we introduce the
Hubble-normalized variables
\be
x=\frac{\dot{\phi}}{\sqrt{6}H}\qquad y= \frac{V}{3H^2}\qquad
\Omega_{\lambda}= \frac{\rho_{T}^2}{6\lambda H^2},\label{3}
\ee
the new temporal variable $\tau=\int H{\rm d}t$, and the
additional dynamical (non-compact) variable, $s$, given by
\be
s=-\partial_{\phi} \ln\;V(\phi).\label{4}
\ee
which is a function of the scalar field.

For the scalar potential treatment, we
proceed following the reference \cite{Fang:2008fw}. Let be defined
the scalar function \be f=\Gamma-1,\qquad \Gamma=
\frac{V^{''}V}{V^{'2}}\label{function}.\ee Since $\Gamma$ is a
function of the scalar field $\Gamma(\phi)$  (see definition
\eqref{function}), also is the variable $s=S(\phi).$ Assuming that
the inverse of $S$ exists, we have  $\phi=S^{-1}(s).$ Thus, one
can obtain de relation $\Gamma=\Gamma(S^{-1}(s))$ and finally the
scalar field potential can be parameterized by a function $f(s).$
Thus, in general, it is possible the treatment of general classes
of potentials using an ``$f$-deviser''. In the table \ref{tab1}
are shown the functions $f(s)$ for some usual quintessence
potentials. The  cases $(a)$-$(c)$ have been studied in $RS2$
branes in \cite{Leyva2009}.
\begin{table*}[tbp]\caption[crit]{Explicit forms of $f(s)$ for some self-interaction potentials.
To homogenize the notations we use units in which $\kappa^2\equiv
8\pi G=1.$}
\begin{center}\begin{tabular}{@{\hspace{4pt}}c@{\hspace{4pt}}c@{\hspace{14pt}}c@{\hspace{14pt}}c}
\hline
\hline\\[-0.3cm]
Label & Potential& $f(s)$& Reference\\[0.1cm]
\hline\\[-0.2cm]
%%%%%%%%
$(a)$ & $V=V_{0}\sinh^{-\alpha}\chi\phi$& $\frac{1}{\alpha}- \frac{\alpha \chi^2}{s^2}$& \cite{Sahni2000, Urena-Lopez2000} \\[0.2cm]
$(b)$ & $V=V_{0}[\cosh(\chi\phi)-1]^{\alpha}$ & $-\frac{1}{2\alpha}+\frac{\alpha\chi^{2}}{2s^{2}}$ & \cite{Sahni2000a}\\[0.2cm]
$(c)$ & $V=\frac{V_{0}}{\left(\eta+e^{-\alpha\phi}\right)^{\beta}}$& $\frac{1}{\beta}+\frac{\alpha}{s}$ & \cite{Zhou:2007xp}\\[0.2cm]
$(d)$ &$V=V_{0}e^{-\chi\phi}+\Lambda$& $-1- \frac{\chi}{s}$& \cite{Cardenas2003}\\[0.2cm]
$(e)$ &$V=V_{0}\frac{e^{\chi\phi^2}}{\phi^m}$& $\frac{s^{2}+8m\chi+s\sqrt{s^{2}+8m\chi}}{2ms^{2}}$& \cite{Brax1999b, Brax2000} \\[0.2cm]
$(f)$ &$V=V_{0}\left[e^{\alpha\phi}+e^{\beta\phi}\right]$& $- \frac{(s+\alpha)(s+\beta)}{s^2}$& \cite{Barreiro2000}\\[0.4cm]
\hline \hline
\end{tabular}\end{center}\label{tab1}
\end{table*}

Bearing this in mind, and using the variables \eqref{3}-\eqref{4}
we deduce the following autonomous system of ordinary differential
equations (ODE)
\begin{align}&x'= \sqrt{\frac{3}{2}}s y-3x+ \frac{3(\Omega_{\lambda}+1)(\gamma-2)}{2(\Omega_{\lambda}-1)}x^{3} + \nonumber \\ & +\frac{3\gamma(\Omega_{\lambda}+1)(y+\Omega_{\lambda}-1)}{2(\Omega_{\lambda}-1)}x \label{eqx}\\
&y'=\frac{3y(\Omega_{\lambda}+1)}{(\Omega_{\lambda}-1)}\left[\left(\gamma-2\right)x^{2}+\gamma\left(\Omega_{\lambda}+y-1\right)\right] -\sqrt{6}xys \label{eqy}\\
&\Omega_{\lambda}'=3\Omega_{\lambda}\left[\left(\gamma-2\right)x^{2}+\gamma\left(\Omega_{\lambda}+y-1\right)\right]\label{equ}\\
&s'=-\sqrt{6}xs^{2}f(s)\label{eqs},\end{align} where the comma
denotes derivative with respect $\tau.$

From the Friedmann equation (\ref{EqF}) follows the relation \be
\Omega_{m}=1-x^{2}-y-\Omega_{\lambda} \label{2.1} \ee Using
\eqref{2.1}, the energy condition $0\leq\Omega_{m}\leq1$ can be
written as \be 0\le x^{2}+y+\Omega_{\lambda}\leq1 \label{2.2}.
\ee

From the definition of $\Omega_{\lambda}$ and the Friedmann
equation we obtain the useful relation
\be%\label{util}
\frac{\rho_T}{\lambda}=
\frac{2\Omega_{\lambda}}{1-\Omega_{\lambda}}\label{OmegaL} \ee

From \eqref{2.4} follows that the invariant set
$\Omega_{\lambda}=1$ corresponds to cosmological solutions where
$\rho_T \gg \lambda$ (corresponding to the formal limit
$\lambda\rightarrow 0$). Therefore, they are associate to high
energy regions, i.e., to cosmological solutions in a neighborhood
of the initial singularity \footnote{See the references
\cite{Foster1998f,Leon2009} for a classical treatment of
cosmological solutions near the initial singularity.}. Due to its
classic nature, our model is not appropriate to describing the
dynamics near the initial singularity, where quantum effects
appear. However, from the mathematical viewpoint, this region
($\Omega_{\lambda}= 1$) is reached asymptotically. In fact, as
some numerical integrations corroborate, there exists an open set
of orbits in the phase interior that tends to the boundary
$\Omega_{\lambda}= 1$ as $\tau\rightarrow-\infty$. Therefore, for
mathematical motivations it is common to attach the boundary
$\Omega_{\lambda}=1$ to the phase space. On the other hand the
points with ($\Omega_{\lambda}= 0$) are associated to the standard
4D behavior ($\rho_T \ll \lambda$ or $\lambda\rightarrow\infty$)
and corresponds to the low energy regime.

From definition (\ref{3}) and from the restriction \eqref{2.2},
and taking into account the previous statements, it is enough to
investigate to the flow of \eqref{eqx}-\eqref{eqs} defined in the
phase space
\begin{align} &\Psi=\{(x,y,\Omega_{\lambda}): 0\leq x^{2}+y+\Omega_{\lambda}\leq1,-1\leq x\leq1,\nonumber\\
& 0\leq y\leq1,0\leq \Omega_{\lambda}\leq 1
\}\times\left\{s\in\mathbb{R}\right\}.\label{2.3}\end{align}

Some cosmological parameters like the equation of state parameter
of the scalar matter $\omega_{\phi}=
\frac{p_{\phi}}{\rho_{\phi}}$, the deceleration parameter
$q=-\left(1+ \frac{\dot{H}}{H^2}\right)$ can be re-expressed as
functions of the new variables as follows \be \omega_{\phi}=
\frac{x^2 -y}{x^2 +y}, \qquad \Omega_{\phi}=x^2 +y \label{2.4} \ee
\be
q=\left(\frac{1+\Omega_{\lambda}}{1-\Omega_{\lambda}}\right)\left[3x^2
+ \frac{3\gamma}{2}\left(1-x^2
-y-\Omega_{\lambda}\right)\right]-1\label{2.4.1} \ee

\subsection{Critical points}

The system \eqref{eqx}-\eqref{eqs}, admits the curves of critical
points $P_1,P_2, P_3$; the critical points $P_4^\pm$ and $P_5;$
and the classes of critical points $P_6^\pm,$ $P_7$ and $P_8$
parameterized by $s^*$ satisfying $f (s^*)= 0.$  In Table
\ref{Point} are displayed the location, existence conditions and
some basic observables of these critical points \footnote{Strictly
speaking the system admits one more curve of critical point with
coordinates $x\in\left[-1,1\right], y=-\frac{x^2 (\gamma
-2)}{\gamma },\Omega_\lambda=1, s=0$, but since the energy
condition \eqref{2.2} is not satisfied, we omit it from the
analysis.}.  The critical points of $P_1$ to $P^{\pm}_6$ always
exist; the point $P_7$ exists for $ s^{*2}\geq3\gamma$, whereas,
$P_8$ exists for $s^{*2}\leq 6$ with $f (s^*)= 0.$

\begin{table*}[tbp]\caption[crit]{Location, existence conditions and some basic observables for the critical points of the system of equations (\ref{eqx})-(\ref{eqs}).}
\begin{center}\begin{tabular}{@{\hspace{4pt}}c@{\hspace{14pt}}c@{\hspace{14pt}}c@{\hspace{14pt}}c@{\hspace{14pt}}c@{\hspace{14pt}}c@{\hspace{14pt}}c@{\hspace{14pt}}c@{\hspace{14pt}}c}
\hline
\hline\\[-0.3cm]
$P_i$ &$x$&$y$&$\Omega_{\lambda}$&$s$& Existence& $\omega_\phi$& $\Omega_\phi$& $q$\\[0.1cm]
\hline\\[-0.2cm]
%%%%%%%%
$P_{1}$& $0$& $1-\Omega_{\lambda}$& $\Omega_{\lambda}\in \left[\right. 0,1\left[\right.$& $0$& \small Always & $-1$& $1-\Omega_{\lambda}$& $-1$\\[0.2cm]
$P_2$& $0$& $0$& $0$& $s\in\mathbb{R}$& "& \small undefined & $0$& $\frac{3\gamma}{2}-1$\\[0.2cm]
$P_{3}$& $0$& $0$& $1$& $s\in\mathbb{R}$& "& \small undefined & $0$& \small undefined\\[0.2cm]
$P_{4}^\pm$& $\pm1$&$0$& $0$& $0$& "& $1$& $1$&$2$\\[0.2cm]
$P_{5}$& $0$&$1$& $0$& $0$& "& $-1$& $1$&$-1$\\[0.2cm]
$P_{6}^\pm$& $\pm1$&$0$& $0$& $s^{*}$& "& $1$& $1$&$2$\\[0.2cm]
$P_{7}$& $\frac{\sqrt{\frac{3}{2}} \gamma }{s^*}$& $-\frac{3 (\gamma -2) \gamma }{2 \left(s^*\right)^2}$& $0$& $s^{*}$& $ s^{*2}\geq3\gamma$& $\gamma-1$& $\frac{3\gamma}{s^{*2}}$&$\frac{3\gamma}{2}-1$\\[0.2cm]
$P_{8}$& $\frac{s^*}{\sqrt{6}}$& $1-\frac{\left(s^*\right)^2}{6}$& $0$& $s^*$& $s^{*2}\leq 6$& $\frac{1}{3} \left(s^{*2}-3\right)$& $1$&$\frac{s^{*2}}{2}-1$\\[0.4cm]
\hline \hline
\end{tabular}\end{center}\label{Point}

\end{table*}

Now let us comment on the stability of the first order
perturbations of (\ref{eqx})-(\ref{eqs}) near the critical points
displayed in table \ref{Point}. Let us comment briefly in their
physical interpretation.

The line of critical points $y=1-\Omega_{\lambda}$  called $P_1$
represent solutions with 5D-corrections, since, in general,
$\Omega_{\lambda}\neq0$. From the relationship between $y$ and
$\Omega_{\lambda}$ follows that this solution is dominated by the
potential energy of the scalar field $\rho_T= V(\phi);$ that is,
it is de Sitter-like  solution ($\omega_\phi=- 1$). In this case
the Friedmann equation can be expressed as \be 3H^2 =V\left(1+
\frac{V}{2\lambda}\right)\label{2.5} \ee In the early universe,
where $\lambda\ll V,$ the expansion rate of the universe for the
RS model differs from the general relativity predictions
\be\frac{H_{RS}}{H_{GR}}= \sqrt{\frac{V}{2\lambda}}\label{2.6} \ee
$P_1$ admits a 2D stable manifold, $M_2$. Due the importance of de
Sitter solutions in the cosmological context, in section
\ref{CMP1} we  calculate explicitly  their center manifold proving
that this critical point is locally asymptotically unstable.

The point $P_2$ represents a matter-dominated solution ($\Omega_m=
1$). Although it is non-hyperbolic, it behaves like a saddle point
 in the space of phase of the RS
model, since they have a nonempty stable and unstable manifolds
(see the table \ref{eigenvalues}) \footnote{Strictly speaking, the concept of saddle point  is not applicable to nonhyperbolic critical points.}.

The critical point $P_3$ is located at the boundary
$\Omega_{\lambda}= 1$ of the phase space region (\ref{2.3}). From
the physical viewpoint, this solution represents the  Big Bang
singularity ($\rho_T\rightarrow\infty$). The eigenvalues for $P_3$
are displayed in tables \ref{eigenvalues}. They were calculated
for orbits contained completely in the invariant set $x=y=0$ and
by taking the limit as $\Omega_\lambda\rightarrow 1.$ For orbits
outside the above invariant set we cannot make the above limit
process since the system is not of class $C^1$ at
$\Omega_\lambda=1$. However, several numerical integrations
suggest that this solution is, indeed, the past attractor.

The critical points $P^{\pm}_4$ are solutions dominated by the
kinetic energy of the scalar field and they represent solutions
with an  standard behavior ($\Omega_{\lambda} = 0$). This critical
points are nonhyperbolic.However, they behave as saddle-like
points in the space of phase because of the instability in the
eigendirection associated with a positive eigenvalue and the
stability of an eigendirection associated to a negative
eigenvalue.

The critical point $P_5$ is a particular case of $P_1$ when
($\Omega_{\lambda} = 0$). They represent a solution dominated by
the potential energy of the scalar field. Indeed,  it is a late
attractor of Sitter provided $f(0)>0$. \footnote{We have arrived
to this conclusion by making the stability analysis of its center
manifold (see section \ref{p5} for an explicit computation).}

The stability analysis of the critical points $P^{\pm}_6$, $P_7$
and $P_8$ is a little more complicated task since the eigenvalues
of the linearization matrix do depend on the function $f(s),$
their zeros, $s=s^*$, and the value of the first derivative at
$s=s^*.$

The critical points $P^{\pm}_6$ are solutions dominated by the
kinetic energy of the scalar field and represent transient states
(saddle points in the phase space) in the evolution of the
universe for $\gamma<2$. For $\gamma=2,$ the point $P^{+}_{6}$ is
nonhyperbolic; the stable manifold is 3D provided \be
s^*>\sqrt{6}\label{6.1} \ee and $f'(s^*)>0;$ otherwise, the stable
manifold it is of dimension less than 3. Similarly, for
$\gamma=2,$ the point $P^{-}_{6}$ is nonhyperbolic;  and their
stable manifold is 3D provided \be s^* <-\sqrt{6}\label{6.2} \ee
and $f'(s^*)<0$; otherwise,  the stable manifold is of dimension
less than 3.
\begin{table*}[tbp]\caption[crit]{Eigenvalues for the critical points of the equations system (\ref{eqx})-(\ref{eqs}).
We use the notation $\beta_\pm=\frac{3}{4}\left(\gamma -2\pm\sqrt{(2-\gamma) \left(\frac{24 \gamma
^2}{\left(s^*\right)^2}-9
   \gamma +2\right)}\right).$}
\begin{center}{\small\begin{tabular}{@{\hspace{4pt}}c@{\hspace{14pt}}c@{\hspace{14pt}}c@{\hspace{14pt}}c@{\hspace{14pt}}c@{\hspace{14pt}}c@{\hspace{14pt}}c}
\hline
\hline\\[-0.3cm]
$P_i$ &$\lambda_1$& $\lambda_2$& $\lambda_3$& $\lambda_4$\\[0.1cm]
\hline\\[-0.2cm]
%%%%%%%%
$P_{1}$& $0$& $0$& $-3$& $-3\gamma$\\[0.2cm]
$P_{2}$& $0$& $-3\gamma$& $3\gamma$& $\frac{3}{2}(\gamma-2)$\\[0.2cm]
$P_{3}$& $0$& $3(\gamma-1)$& $3\gamma$& $6\gamma$& \\[0.2cm]
$P_{4}^\pm$& $0$& $-6$& $6$& $6-3\gamma$\\[0.2cm]
$P_{5}$& $0$& $0$& $-3$& $-3\gamma$& \\[0.2cm]
$P_{6}^\pm$& $-6$& $6-3\gamma$& $6\mp\sqrt{6}s^{*}$& \small $\mp\sqrt{6}(s^{*})^2 f'(s^{*})$\\[0.2cm]
$P_{7}$& $-3\gamma$& $\beta_-$& $\beta_+$& $-3\gamma s^{*}f'(s^{*})$\\[0.2cm]
$P_{8}$& $\frac{1}{2}\left(s^{*2}-6\right)$& $s^{*2}-3\gamma $& $-s^{*2}$& $-s^{*3}f'(s^{*})$\\[0.4cm]\hline \hline
\end{tabular}}\end{center}\label{eigenvalues}
\end{table*}

The critical points $P_7$ are nonhyperbolic for
$s^*\in\left\{-\sqrt{3\gamma},\sqrt{3\gamma}\right\}$ or $s^*
f'(s^*)=0$ or $\gamma=2.$ The points $P_8$ represent scalar-field-dominated solutions ($\Omega_\phi= 1$) which are non hyperbolic provided
${s^*}^2\in \{0, 3\gamma, 6\}$ or  $f'(s^*)=0.$

Having presented the eigenvalues of the Jacobian matrix for the
critical points $P_7$ and $P_8$ in table \ref{eigenvalues}, we straightway formulate the following
results.

The sufficient conditions for the asymptotic stability of
the matter-scalar-field scaling solution ($P_7$) are
    \begin{enumerate}
   \item[i)] $0\leq\gamma< 2,\, s^*<-\sqrt{3\gamma}$ and
$f'(s^*)<0,$ or \item[ii)] $0\leq\gamma\leq 2,\, s^*>\sqrt{3\gamma}$
and $f'(s^*)>0.$
\end{enumerate}
The sufficient conditions for the asymptotic stability of the
scalar-field-dominated solution ($P_8$) are either
   \begin{enumerate}
   \item[iii)] $0\leq\gamma< 2,\, -\sqrt{3\gamma}<s^*<0$ and
$f'(s^*)<0,$ or \item[iv)] $0\leq\gamma\leq 2,\, 0<s^*<\sqrt{3\gamma}$
and $f'(s^*)>0.$
   \end{enumerate}

\subsubsection{Dynamics of the center manifold of $P_5$}\label{p5}
The solution $P_5$ is a particular case of $P_1,$ which can be a
candidate to be a late-time de Sitter attractor without
5D-corrections ($\Omega_{\lambda}= 0$). To analyze its stability we carry out a detailed stability study of their center manifold using the Center Manifold Theory \cite{leon2011a}.

Introducing the new variables \be x_1=s,\,
x_2=\Omega_\lambda,\,y_1=x-\frac{s}{\sqrt{6}},\,y_2=y+\Omega_\lambda
   -1\label{vvar2},\ee and Taylor expanding the evolution equations for the new variables \eqref{vvar2}, we obtain the vector field
\be x_{1}'= -x_1^2 \left(x_1+\sqrt{6} y_1\right)
f(0)+\mathcal{O}(4),\label{eqx1} \ee \be  x_{2}'=\frac{1}{2} x_2
\left(x_1^2+2 \sqrt{6} y_1
   x_1+6 y_1^2\right) (\gamma -2)+3 x_2 y_2\gamma +\mathcal{O}(4),\label{eqx2} \ee
\begin{align} & y_{1}'=-3 y_1+\frac{1}{4} \left(-\sqrt{6} x_1 (2
   x_2+y_2 (\gamma -2))-6 y_1 y_2 \gamma \right)+\nonumber\\ &+\frac{1}{24} \left(\sqrt{6} (-\gamma +4 f(0)+2)
   x_1^3+6 y_1 (-3 \gamma +4 f(0)+6) x_1^2+\right. \nonumber \\ & \left.-6 \sqrt{6} \left(3 (\gamma -2) y_1^2+2 x_2 y_2
   \gamma \right) x_1+\right. \nonumber \\ & \left. -36 \left((\gamma -2) y_1^3+2 x_2 y_2 \gamma  y_1\right)\right)+\mathcal{O}(4),\label{eqy1}
   \end{align}
and \begin{align}  & y_{2}'=-\frac{\gamma
   x_1^2}{2}+\left(x_2-\frac{y_2 \gamma }{2}\right) x_1^2-\sqrt{6} y_1 (\gamma -1)
   x_1\nonumber \\ &+\sqrt{6} y_1 (x_2+y_2-y_2 \gamma ) x_1-3 y_2 \gamma +\nonumber\\ & -3 y_2 \left((\gamma
   -2) y_1^2+2 x_2 y_2 \gamma \right)+\nonumber\\ &-3 \left((\gamma -2) y_1^2+y_2^2 \gamma \right)+\mathcal{O}(4),\label{eqy2}
   \end{align}
   where $\mathcal{O}(4)$ denotes error terms of fourth order in
   the vector norm.

Accordingly  to the Center Manifold theorem, the local center
manifold of the origin for the vector field
\eqref{eqx1}-\eqref{eqy2} is given by the graph \begin{align} &
W_{\text{loc}}^c(\mathbf{0})=\left\{(x_1,x_2,y_1,y_2):
y_1=F(x_1,x_2), \right. \nonumber\\ & \left. y_2=G(x_1,x_2),
x_1^2+x_2^2<\delta\right\}\label{func}\end{align} where $\delta>0$ is a
small enough real value.

Deriving each one of the functions in (\ref{func}) with respect
$\tau$ one can obtain the system of quasi-lineal partial
differential equations \be y_1'-\frac{\partial F}{\partial
x_1}x_1'-\frac{\partial F}{\partial x_2}x_2'=0\label{EDP1} \ee \be
y_2'-\frac{\partial G}{\partial x_1}x_1'-\frac{\partial
G}{\partial x_2}x_2'=0\label{EDP2}.\ee Since we have used Taylor
expansions up to third order for obtaining the system
\eqref{eqx1}-\eqref{eqy2} we must seek a solution for
\eqref{EDP1}-\eqref{EDP2} in the following form (see \cite{coley2003b, Tavakol2005B, leon2011a}): \begin{align} & F(x_1 ,x_2
)= a_1 x_1^3+a_2 x_1^2+a_3 x_1^2+a_4 x_1 x_2^2+a_5 x_1
x_2+\nonumber
\\ &+a_6 x_2^3+a_7 x_2^2+\mathcal{O}(4) \label{solucion1} \end{align}
\begin{align} & G(x_1 ,x_2 )= b_1 x_1^3+b_2 x_1^2+b_3 x_1^2+b_4 x_1
x_2^2+b_5 x_1 x_2+\nonumber \\ &+b_6 x_2^3+b_7
x_2^2+\mathcal{O}(4), \label{solucion2} \end{align} as $x_{i} \rightarrow
0$  where
$\mathcal{O}(4) $ is an error term of fourth order in the vector
norm. Substituting expressions \eqref{solucion1} and
\eqref{solucion2} in the equations \eqref{EDP1}-\eqref{EDP2}, and
comparing terms of equal powers, we obtain that the non-null
coefficients in the above expressions \eqref{solucion1} and
\eqref{solucion2} are \be a_1=\frac{f(0)}{3 \sqrt{6}},
\quad a_5=-\frac{1}{\sqrt{6}}, \quad b_2=- \frac{1}{6}, \quad
b_3=\frac{1}{3},\label{coeficiente} \ee i.e.,
\begin{align} y_1 =F(x_1 ,x_2 )=-\frac{x_1 x_2}{\sqrt{6}}+\frac{x_1^3 f(0)}{3 \sqrt{6}}+\mathcal{O}(4),\nonumber\\
  y_2 =G(x_1 ,x_2 )=-\frac{x_1^2}{6}+\frac{x_1^2
   x_2}{3}+\mathcal{O}(4)\label{y1y2}
\end{align}Thus, the dynamics on the center manifold,  is given by \begin{align}
&x_{1}'= -x_1^3  f(0)+\mathcal{O}(4)\label{eqx1s}\\
&x_{2}'= -x_1^2 x_2 +\mathcal{O}(4)\label{eqx2s}. \end{align}

Neglecting the error terms, and introducing the coordinate
transformation $u_1=x_1^2,$ the system \eqref{eqx1s}-\eqref{eqx2s}
reduces to the simpler form \begin{align} & u_{1}'=-2 u_1^2
f(0)\label{equ1s}\\
&x_{2}'= -u_1 x_2\label{eqxx2s},\end{align} where the region of
physical interest is $u_1\geq 0, x_2\geq 0.$

Observe that the
dynamics on the center manifold, governed by
\eqref{equ1s}-\eqref{eqxx2s}, depends on the value $f(0).$ If
either $f (0)= 0$ or $f$ is singular at the origin, the system
\eqref{equ1s}-\eqref{eqxx2s} does not represents correctly the
dynamics of the center manifold. I such a case, we must
incorporate higher order terms in the scheme, increasing the
problem complexity. Thus, we assume that $f(0)$ is a real number
such that  $f(0)\neq 0.$

\begin{figure}
\begin{center}
\includegraphics[height=3in,width=3in]{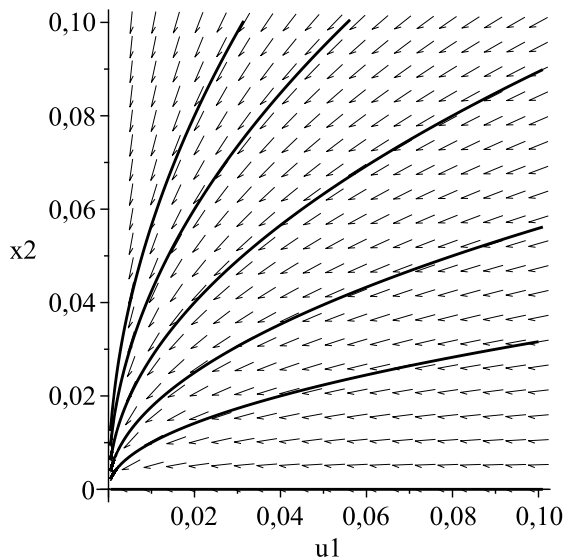}
\begin{center} (a) \end{center}
\bigskip
\includegraphics[height=3in,width=3in]{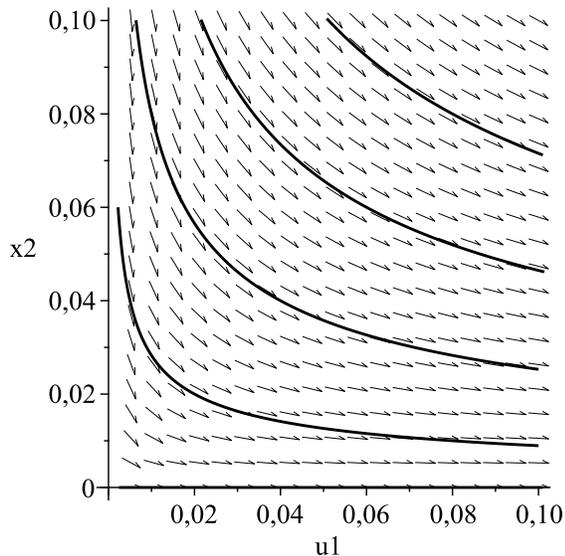}
\begin{center} (b) \end{center}
\caption{\label{f0_pm0_1} Phase space of the system
(\ref{equ1s})(\ref{eqx2s}) for: (a) $f(0)=1$ and  (b)
$f(0)=-0.1.$}
\end{center}
\end{figure}

According to the Center Manifold Theorem, the stability analysis
of $P_5$ is reduced to the analysis of the stability of the origin
of the system \eqref{equ1s}-\eqref{eqxx2s}. For this analysis we
resort to numerical investigation. In figure \ref{f0_pm0_1} are
displayed several orbits contained the physical region $u_1\geq 0,
x_2\geq 0.$ Observe that the axis are invariant sets. For $f(0)>0$
(see the panel (a) in figure \ref{f0_pm0_1}), there is an open
sets of orbits that converge to the origin as time goes forward;
thus, the origin is asymptotic stable for initial conditions in a
vicinity of the origin whenever $f(0)>0.$ From the asymptotic
stability of the origin of (\ref{eqx1s})-(\ref{eqx2s}) follows that, for  $f(0),$ the center manifold
of $P_5$ is locally asymptotic stable, and hence, the solution
$P_5$ of the system(\ref{eqx})-(\ref{eqs}) also is. Therefore,
$P_5$ with $f(0)>0$ corresponds to a late time de Sitter
attractor. This result for RS2 brane cosmology is in a perfect
agreement with the standard 4-dimensional TGR framework.

%%%%%%%%%%%%%%%%%%%%%%%%%%%

\subsubsection{Dynamics of the center manifold of  $P_1$}\label{CMP1}
In this section we investigate the stability of the curve of
critical points $P_1$ for $0<\Omega_{\lambda}<1.$ This solutions
correspond to a de Sitter expansion with 5D-corrections. According
to the RS2 model this solution cannot behave like a late time
attractor since 5D-corrections are typical of the high energy
(early universe) and not for low energy (universe late) regimes. If
we can prove that this solution is of saddle type, we can
correlate this behavior with a transient inflationary stage for
the universe. In order to verify our claim, we appeal to the Center
Manifold Theory.

Let us consider an arbitrary critical point with coordinates
$(x=0, y=1-u_c,\Omega_\lambda=u_c, s=0)$ located at $P_1.$

In order to prepare the system (\ref{eqx})-(\ref{eqs}) for the
application of the Center Manifold Theorem we introduce the
coordinate change \begin{align} & u_1=-\frac{s (u_c-1)}{\sqrt{6}},\,
u_2=-\Omega_\lambda -u_c
   (y+\Omega_\lambda -2),\nonumber \\ & v_1=(u_c+1) (y+\Omega_\lambda -1),\, v_2=\frac{s
   (u_c-1)}{\sqrt{6}}+x\label{Vvar2}.\end{align} Then, we Taylor expand the system $u_1', u_2', v_1', v_2'$
in a neighborhood of the origin with error of order
$\mathcal{O}(4).$

Accordingly  to the Center Manifold theorem, the local center
manifold of the origin for the resulting vector field is given by
the graph: \begin{align} &
W_{\text{loc}}^c(\mathbf{0})=\left\{(u_1,u_2,v_1,v_2):
v_1=F_1(u_1,u_2), \right. \nonumber\\ & \left. v_2=G_1(u_1,u_2),
u_1^2+u_2^2<\delta\right\}\label{Center}\end{align} for $\delta>0$ a
small enough real value.

The functions $F_1$ and $G_1$ in definition \eqref{Center} are a
solution of a system of quasi-linear differential equations
analogous to \eqref{EDP1}-\eqref{EDP2}. This system should be
solved with an error of order $\mathcal{O}(4),$  obtaining the
functional dependence: \begin{align} && v_1=\frac{2 u_1^2 u_2
   (u_c+1)}{u_c-1}-u_1^2 (u_c+1),\nonumber\\
&&v_2=\frac{2
   \left(u_1^3 u_c-u_1^3
   f(0)\right)}{u_c-1}-\frac{u_1
  u_2}{u_c-1}\label{soluciones2}.\end{align}

Then, the dynamics on the center manifold is given by
\begin{align} &&u_1'=\frac{6 u_1^3 f(0)}{u_c-1}+\mathcal{O}(4)\label{sist1} \\
&&u_2'=6 u_c u_1^2+\frac{6 u_2 (1-3 u_c)
   u_1^2}{u_c-1}+\mathcal{O}(4)\label{sist2}.\end{align}

In the same way as for $P_5$, the dynamics of the system
\eqref{sist1}-\eqref{sist2} depends on the values of $f(0).$ We
assume that $f(0)\in\mathbb{R}\setminus \{0\}.$ Otherwise it is
required to include higher order terms in the Taylor expansion,
increasing the numerical complexity.

\begin{figure}
\begin{center}
\includegraphics[height=3in,width=3in]{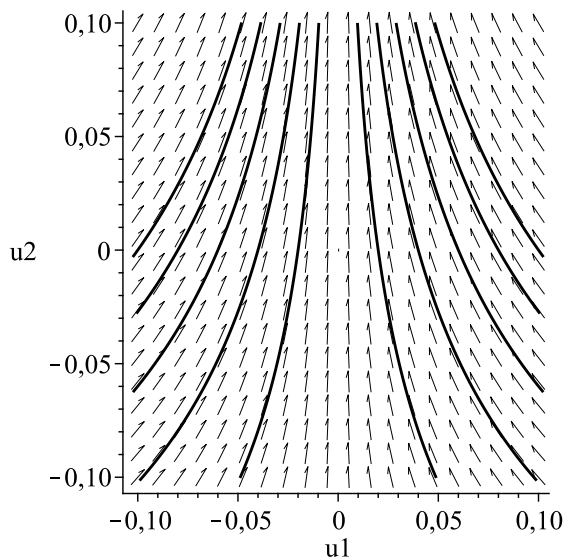}
\begin{center} (a)\end{center}
\bigskip
\includegraphics[height=3in,width=3in]{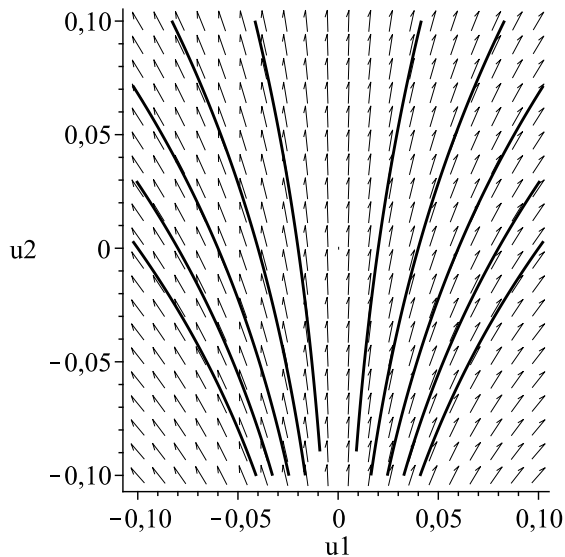}
\begin{center} (b)\end{center}
\caption{\label{VC} Phase space of the system
\eqref{sist1}-\eqref{sist2} for the choices  (a) $f(0)= 2$ and
$u_c= 0.5$ (b) $f(0)=- 2$ and $u_c= 0.5$. Observe that the axis
$u_1$ is a line of points of critical asymptotic unstable in both
cases for initial conditions in a vicinity of the origin. The
origin behaves as a saddle point.}
\end{center}
\end{figure}

In figure \eqref{VC} are displayed some orbits in the phase space of
the system \eqref{sist1}-\eqref{sist2} for the choices:  (a)
$f(0)= 2$ and $u_c= 0.5$  and (b) $f(0)=- 2$ and $u_c= 0.5$. The
origin of coordinates is locally asymptotically unstable (of
saddle type) irrespective the sign of $f(0)$. Henceforth, the
center manifold of $P_1$ is locally asymptotic unstable (saddle
type) for $f(0)\neq0;$ also is $P_1.$

The physical interpretation
of this result is that there are not late time attractors with
5D-modifications. This type of corrections are characteristic of
the early universe. In this sense the cosmological solution
associated to the critical $P_1$ correlates with the primordial
inflation.

%%%%%%%%%%%%%%%%%%%%%%%%%%%

\section{Exponential Potential}\label{sIV}

The objective of this section is to illustrate our analytical
results for the exponential potential, \be
V(\phi)=V_{0}e^{-\chi\phi}+\Lambda\label{exppot}.\ee This
potential have been widely investigated in the literature. It was
studied for quintessence models in \cite{Cardenas2003} were it is
considered a negative cosmological constant $\Lambda$. In our case
of interest we assume $\Lambda\geq 0,$ to avoid dealing with
negative values of $y.$ However we can apply our procedure by
permitting negative values for $y$ for the case $\Lambda< 0.$ The
dark energy models with exponential potential and negative
cosmological constant were baptized as Quinstant Cosmologies. They
were investigated in \cite{Leon2009a} using an alternative
compactification scheme. The asymptotic properties of a
cosmological model with a scalar field with exponential potential
have been investigated in the context of the General Relativity by
the authors of \cite{Fang:2008fw, Copeland1998}, and in the context
of the RS braneworlds  by \cite{Goheer2003, Gonzalez2009}. In both
cases it was studied the pure exponential potential ($\Lambda=
0$). Potentials of exponential orders at infinity were studied in
the context of Scalar-tensor theories and conformal $F(R)$
theories by the authors of \cite{Leon2009, leon2011a}.

We comment
that the procedure introduced in previous sections is fairly
general and can be applied to others potentials as those showed in
table \ref{tab1}. We remain in the exponential potential for
simplicity. Also, we consider a pressureless (dust) background,
i.e., $\gamma=1.$

The function $f(s)$ corresponding to the potential \eqref{exppot}
is given by \be f(s)=-1- \frac{\chi}{s} \label{2.11}. \ee The zero
of this function is \be s^* =-\chi \qquad f'(s^{*})=
\frac{\chi}{s^{*2}}= \frac{1}{\chi}\label{2.12}.\ee

Observe that
for the potential \eqref{exppot} $s^* f'(s^{*})<0.$ Thus, the only
relevant late-time attractor should the de Sitter solution. In
fact, the critical points $P_7$ of the table \ref{Point} are reduced
to the single point \be P_7 =\left(-\sqrt{\frac{3}{2}}\frac{1}{\chi},\,\,
-\frac{3}{2\chi^2} ,\,\\0,\,\, \chi\right) \label{2.13},\ee which
represents a saddle point in the  phase space. The critical points
$P_8$ are reduced to the single one \be P_8 =\left(-\frac{\chi}{\sqrt{6}},\,\, 1-
\frac{\chi^2}{6},\,\,0,\,\,-\chi\right) \label{2.14}.\ee This
point represents a scalar-field dominated solution
($\Omega_{\phi}=1$). It is a saddle point in the phase space.
Observe that all the trajectories in the phase space always
emerge from the point $(x,y,\Omega_{\lambda})=(0,0,1)$.

%%%%%%%%%%%%%%%%%%%%%%%%%%%
\begin{figure}[t]
\begin{center}
\includegraphics[height=2in,width=2in]{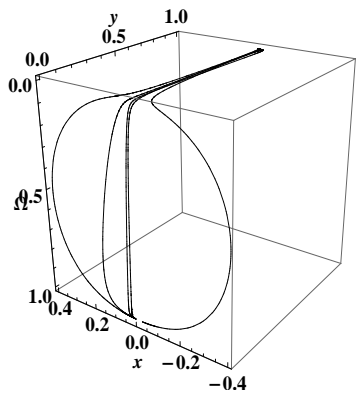}
\begin{center}(a)\end{center}
\bigskip
\includegraphics[height=2in,width=2in]{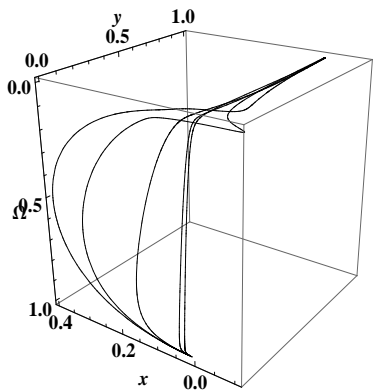}
\begin{center}(a)\end{center}
\vspace{0.2cm}
\bigskip
\caption{Some orbits in the projection $(x,y,\Omega_{\lambda})$ of the phase space for  \eqref{eqx}-\eqref{eqs}) and potential $V(\phi)=V_{0}e^{-\chi\phi}+\Lambda$ for the choice (a) $(\gamma,\chi)=(1,0.5)$, and (b) $(\gamma,\chi)=(1,100)$.}
\label{SD2}
\end{center}
\end{figure}
%%%%%%%%%%%%%%%%%%%%%%%%%%%

In the Fig \ref{SD2}, we present, for different choices of the
free parameters, two numerical integrations which suggest that $P_5$ that is a de Sitter late-time
attractor with an standard 4D behavior  ($\Omega_{\lambda}=0$). However, in order the prove this claim we need to use the Center Manifold Theory. Although for the potential \eqref{exppot} the result in the appendix \label{CMP10} does not apply since  $f(0)=-\text{sgn}(\chi)\infty$ is not a real number, we can use the same procedure as in section  \ref{CMP10} to obtain the center manifold of $P_5$ by setting from the beginning the functional form of $f(s)$ given by \eqref{2.11}.

For  $P_5,$ the graph of the center manifold is given, up to an error term $\mathcal{O}(4)$,   by
\begin{align} & y_1=\frac{\left(\chi ^2-1\right) x_1^3}{3 \sqrt{6}}+\left(\frac{1}{3} \sqrt{\frac{2}{3}} x_2 \chi -\frac{\chi }{3
   \sqrt{6}}\right) x_1^2-\frac{x_2 x_1}{\sqrt{6}},\nonumber\\ & y_2=\frac{\chi  x_1^3}{9}+\left(\frac{x_2}{3}-\frac{1}{6}\right) x_1^2\label{EXPy1y2}
\end{align} where we have introduced the variables \eqref{vvar2}.

The dynamics on the center manifold is governed by the equations
\begin{align} & x_1'=\left(1-\frac{\chi ^2}{3}\right) x_1^3+\chi(1 -x_2) x_1^2 +\mathcal{O}(4)\label{Expp10equ1s},\\& x_2'=-x_1^2 x_2+\mathcal{O}(4)\label{Expp10equ2s},\end{align} defined in the phase plane $x_1\in\mathbb{R},x_2\in[0,1].$

\begin{figure}
\begin{center}
\includegraphics[height=2.5in,width=2.5in]{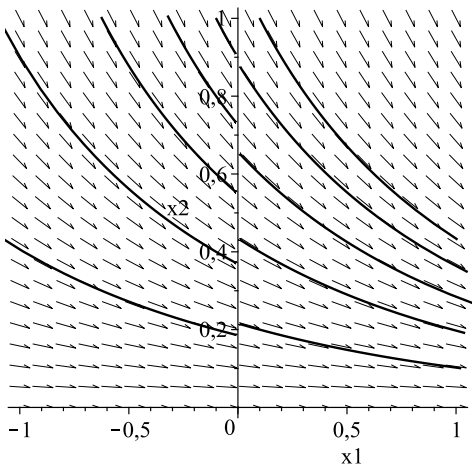}
\begin{center} (a) \end{center}
\includegraphics[height=2.5in,width=2.5in]{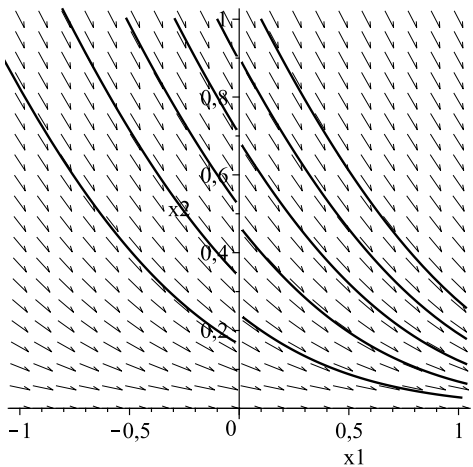}
\begin{center} (b) \end{center}
\caption{\label{chi} Some orbits of the flow of \eqref{Expp10equ1s}-\eqref{Expp10equ2s} for the choices: (a)  $\chi=0.5,$ and  (b) $\chi=-0.5.$  In both cases the $x_2$-axis  is invariant. In (a) the physical region is the semiplane $x_1>0$; for orbits with $s(0)>0$ the origin is unstable to perturbations in the $s$-direction. In (b) the physical region is the semiplane $x_1<0$. Thus for orbits with $s(0)<0$ the origin (hence $P_5$) is asymptotically stable.}
\end{center}
\end{figure}

In general, the system \eqref{eqx}-\eqref{eqs} is not invariant under the change $(s,x)\rightarrow (-s,-x)$ unless $f(s)$ is an even function, i.e., $f(-s)=f(s).$ However, in the particular case of the potential \eqref{exppot} we observe from \eqref{2.11} that the system \eqref{eqx}-\eqref{eqs} is invariant under the discrete symmetry $(s,x,\chi)\rightarrow (-s,-x,-\chi).$ It is easy to show that for the exponential case the symmetry $(s,x,\chi)\rightarrow (-s,-x,-\chi)$ is induced by the discrete symmetry $(\phi, \chi)\rightarrow (-\phi, -\chi)$.  On the other hand, for the potential \eqref{exppot} the function $s(\phi)$ is given by
\begin{equation*}s(\phi)=\frac{V_0 \chi }{V_0+e^{\phi  \chi } \Lambda }.\end{equation*} Assuming $V_0>0$ and $\Lambda>0$ we have $\text{sgn}(s)=\text{sgn}(\chi).$ As we see, the function $s(\phi)$ is not defined for all the real values, that is, the region of physical interest is restricted to one semiplane $s \lessgtr 0$ (or in the degenerate case $\chi=0$ to the hyperplane $s=0$) depending of the sign of $\chi$. That is, $s$ is either positive, negative or zero, for $\chi$ either positive, negative or zero, provided $V_0>0$ and $\Lambda>0.$ This fact limits the application of the above discrete symmetry. In the figures \eqref{chi} (a) and \eqref{chi} (b) are displayed some orbits of the flow of \eqref{Expp10equ1s}-\eqref{Expp10equ2s} for the choice $\chi=+0.5,$ and $\chi=-0.5,$ respectively. In both cases, the $x_2$-axis is invariant so the orbits cannot cross throught it, i.e., the orbits with $s(0)\lessgtr 0$ at the intial time, remains in the respective semiplane all the time. Also, in both figures the flow is the same, thus, the discrete symmetry $(s,x,\chi)\rightarrow (-s,-x,-\chi)$ is preserved. However, from the physical view point, for the choice $\chi=+0.5$ the physical region is the right semiplane ($x_1\equiv s>0$). Thus, the physical orbits depart form the origin as the time goes forward. This means that the origin is unstable. For the choice $\chi=-0.5,$ the physical region is the left semiplane $x_1\equiv s<0,$ thus, the physical orbits approach the origin as the time goes forward.

Summarizing, for $V_0>0$ and $\Lambda>0$ we have $\text{sgn}(s)=\text{sgn}(\chi).$ Thus, for the potentials \eqref{exppot} with $\chi<0$ the de Sitter solution ($P_5$) is asymptotically stable. This fact is self-consistent with the condition $f(0)=-\text{sgn}(\chi)\infty \gg 0$; the sufficient condition for the stability of the de Sitter solution derived in section \ref{p5}. However, for the potentials \eqref{exppot} with $\chi>0$ the de Sitter solution is unstable (of saddle type) to perturbations in the $s$-direction. In this case the late-time solution is given by the asymptotic configuration $x=0, y=1, \Omega_\lambda=0, \phi\rightarrow -\infty.$

\section{Results and discussion}\label{sV}

The main results of this paper can be summarized as follows. The
critical point $P_3=(0, 0, 1)$ represents a Big Bang singularity.
According to our numerical integrations in Figs. \ref{SD2}, we
observe that all the trajectories in the phase space, but a
measure zero set, emerge from the vicinity of this point. This
result agrees with the previous results obtained in
\cite{Leyva2009}. In \cite{Leyva2009} the authors use
another coordinate system, that is equivalent, except
diffeomorphisms, to the system used in this paper.

In the
particular case of a scalar field with potential
$V=V_{0}e^{-\chi\phi}+\Lambda$ trapped in the brane, we have
proved that for $\chi<0$ the de Sitter solution ($P_5$) is asymptotically stable. However, for $\chi>0$ the origin, i.e., $P_5$ is unstable (of saddle type) and the late-time solution is given by the asymptotic configuration $x=0, y=1, \Omega_\lambda=0, \phi\rightarrow -\infty.$ This class of potentials contains the previously
studied potentials in \cite{Gonzalez2009} with $\Lambda= 0$. Thus our present results generalize those in \cite{Gonzalez2009}.

In the general case, for potentials satisfying $f(0)\in\mathbb{R},$ we have the following results. By an explicit computation of the center manifold of $P_1$ and of
$P_5$ we prove that
\begin{itemize}
\item $P_1$ is locally asymptotic unstable (of saddle type)
irrespectively the sign of $f(0)\in\mathbb{R}\setminus \{0\}.$
This feature is corroborated in the Figures \ref{VC}. \item $P_5$
is locally asymptotically stable for  $f(0)>0$ and unstable (of
saddle type) for $f(0)<0.$ This result is illustrated in the
Figures \ref{f0_pm0_1} and \ref{chi}.
\end{itemize}

The solutions dominated by the kinetic energy of the scalar field
$P_4^{\pm}$ and $P_6^{\pm}$ behave like  saddle-type solutions.
This is a main difference with respect the standard 4D theory
where this type of solutions are always past attractors.

In this general case, the possible late-time attractors are:
\begin{itemize} \item the standard 4D de Sitter solution $P_5$
($\omega_{\phi}=-1$) whenever $f(0)>0$; \item the
matter-scalar-field scaling solution $P_7$ ($\Omega_{\phi}\sim
\Omega_{m}$). The sufficient conditions for its asymptotic
stability are $s^*<-\sqrt{3\gamma}, f'\left(s^*\right)<0$ or
$s^*>\sqrt{3\gamma}, f'\left(s^*\right)>0;$ and \item the
scalar-field-dominated solution $P_8$ ($\Omega_{\phi}=1$). The
sufficient conditions for its asymptotic stability are
$-\sqrt{3\gamma}<s^*<0, f'\left(s^*\right)<0$ or
$0<s^*<\sqrt{3\gamma}, f'\left(s^*\right)>0.$ \end{itemize}

\section{Conclusions}\label{sVI}
In the present paper we have investigated the phase space of the
Randall-Sundrum braneworlds models with a self-interacting scalar
field trapped in the brane with arbitrary potential.

From our
numerical experiments we claim that $P_3$ is associated with the
Big Bang singularity type. The numerical investigation suggest
that it is always the past attractor in the phase space of the
Randall-Sundrum cosmological models.

Using the center manifold theory we have obtained sufficient conditions for the
asymptotic stability of de Sitter solution.

We have obtained
conditions on the potential for the stability of the scaling
solutions as well for the stability of the scalar-field dominated
solution.

We have proved, using the center manifold theory and
numerical investigation, that there are not late time attractors
with 5D-modifications since they are always saddle-like. This fact
correlates with a transient primordial inflation.

In the
particular case of a scalar field with potential
$V=V_{0}e^{-\chi\phi}+\Lambda$ we have proved  that for $\chi<0$ the de Sitter solution is asymptotically stable. However, for $\chi>0$ the de Sitter solution is unstable (of saddle type).

\section*{Acknowledgments}
This work was partially supported by PROMEP, DAIP, and by CONACyT, M\'exico, under grant 167335; by MECESUP FSM0806, from Ministerio de Educaci\'on de Chile; and by the National Basic Science Program (PNCB) and Territorial CITMA Project (No. 1115), Cuba. DE, CRF and GL wish to thank the MES of Cuba
for partial financial support of this investigation. YL is grateful to the Departamento de F\'isica and the
CA de Gravitaci\'on
y F\'isica Matem\'atica for their kind hospitality and their joint support for a
postdoctoral fellowship. The authors wish
to thank two anonymous referees for their useful comments and for bringing our attention to
several references.

%\bibliographystyle{unsrt}
%\bibliography{bib}

\end{document}